\documentclass[twocolumn,usenatbib]{mnras}

\usepackage{graphicx}
\usepackage{amsmath}

\def\pmb#1{\setbox0=\hbox{#1}%
    \kern-.025em\copy0\kern-\wd0
    \kern.05em\copy0\kern-\wd0
    \kern-.025em\raise.0433em\box0}

\def\ltsima{$\; \buildrel < \over \sim \;$}
\def\gtsima{$\; \buildrel > \over \sim \;$}
\def\simlt{\lower.5ex\hbox{\ltsima}}
\def\simgt{\lower.5ex\hbox{\gtsima}}
\def\p2Y{\;_2Y}
\def\m2Y{\;_{-2}Y}
\def\bs{\boldsymbol}

\def\mk2{\mu {\rm K}^2}
\def\Planck{\it Planck\rm}

\def\LCDM{$\Lambda$CDM}

\newcommand{\Mpc}{\text{Mpc}} 
\newcommand{\Hunit}{~\text{km}~\text{s}^{-1} \Mpc^{-1}}

\def\pmb#1{\setbox0=\hbox{#1}%
     \kern-.025em\copy0\kern-\wd0
     \kern.05em\copy0\kern-\wd0
     \kern-.025em\raise.0433em\box0}

\begin{document}

\title[Use of $H_0$ priors]{To $H_0$ or not to $H_0$?}

\author[George Efstathiou]{George Efstathiou\\
 Kavli Institute for Cosmology Cambridge and 
Institute of Astronomy, Madingley Road, Cambridge, CB3 OHA.}

\maketitle

\begin{abstract} 
This paper investigates whether changes to late time physics can resolve the `Hubble tension'.
It is argued that many of the claims in the literature favouring such solutions are caused by a 
misunderstanding of how distance ladder measurements actually work and, in particular, by the inappropriate
use of a distance ladder $H_0$ prior. A dynamics-free inverse distance ladder shows that changes
to late time physics are strongly constrained observationally and 
cannot resolve the discrepancy between the SH0ES data and the base \LCDM\
cosmology inferred from \Planck. We propose a statistically rigorous scheme to replace the use
of $H_0$ priors.

\end{abstract}


\section{Introduction}
\label{sec:Introduction}

As is well known, a six parameter \LCDM\ cosmology\footnote{Which I
  will refer to as the base \LCDM\ model.}  has proved to be
spectacularly successful in explaining the cosmic microwave background
radiation (CMB), light element  abundances and a wide range
of other astronomical data \citep[e.g.][]{Params:2018,
  Efstathiou:2019, Mossa:2020,Alam:2020}. As noted in
\cite{Params:2018}, the agreement between the base \LCDM\ model and
observations is so good, that many researchers have begun to focus on
possible discrepancies or `tensions', with the hope that the model
might break to reveal new truths about our Universe. This is reasonable given that
 many ingredients of the model, particularly the physics describing the dark sector, remain 
mysterious at this time.

The discrepancy between early time and late time determinations of the
Hubble constant, $H_0$, is probably the most serious such
tension. This tension  became apparent following the first results from
the \Planck\ satellite \citep{Params:2014} which revealed a
discrepancy between the best fit base \LCDM\ value of $H_0$ and the
Cepheid-based distance ladder measurement of $H_0$ by the
SH0ES\footnote{SNe, $H_0$, for the Equation of State of dark energy}
collaboration \citep{Riess:2011}. Since then, the `Hubble tension' (as
it has become known) has intensified: recent results from the SH0ES
collaboration give $H_0 = 74.03 \pm 1.42 \Hunit$ \citep[][hereafter R19]{Riess:2019}
(udpating the results of \cite{Riess:2016}, hereafter R16) which differs by $4.3\sigma$ from the base \LCDM\
value $H_0 = 67.44
\pm 0.58 \Hunit$ inferred from the most recent analysis of
\Planck\ \citep{Efstathiou:2019}. To add to the conundrum, the lower value of $H_0$ inferred
from the CMB is in very good agreement with various applications of an inverse
distance ladder, irrespective of whether the sound horizon, $r_d$,
is fixed to a value determined from the CMB or to a value inferred from primordial nucleosynthesis
\citep[e.g.][]{Aubourg:2015, Verde:2017, Addison:2018, Abbott:2018,
  Macaulay:2019}.

Possible modifications to the base \LCDM\ model that might resolve
this tension have been discussed in the reviews by \cite{Knox:2020},
\cite{Beenakker:2021} and \cite{diValentino:2021}. In broad brush, the
proposed solutions fall into four categories: (i) radical departures
from conventional cosmology, including departures from General
Relativity\footnote{Such models, including those leading to local changes in the properties of Type Ia
SN \citep{Alestas:2021}  and/or Cepheids \citep{Desmond:2019},  will not be considered further in this paper.}; 
(ii) changes to the physics of the early Universe (for
example adding additional relativistic species, or neutrino
interactions); (iii) new physics at matter-radiation equality, or
recombination,  that alters the value of the sound horizon, (iv) changes to the
expansion history at late times. The focus of this paper is on solutions in 
class (iv).

The issue of whether the tension is real 
is not yet fully clear \citep[see][]{Freedman:2019, Yuan:2019,Freedman:2020}.
Despite the fact that the author is an unashamed  Hubble tension skeptic \citep{Efstathiou:2020},  
 I will take the SH0ES results at face value in this paper and consider whether the 
Hubble tension can be resolved by modifications to the \LCDM\  late time expansion history.

\begin{figure}
	\centering
	\includegraphics[width=85mm, angle=0]{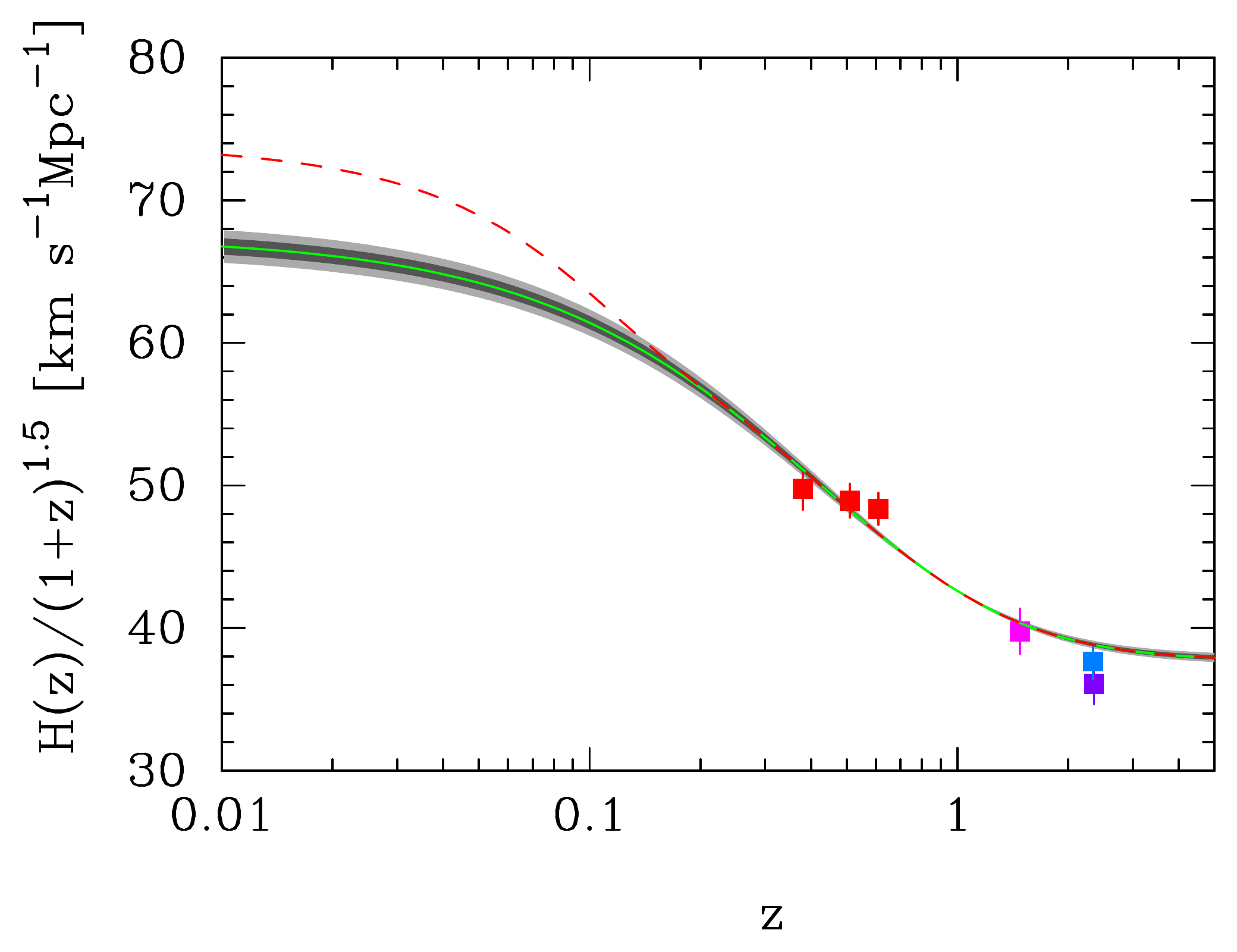} 
	\caption{The evolution of the Hubble parameter with
          redshift. The red points\protect\footnotemark \ \ show $H(z)$
          measurements in three redshift bins inferred from galaxy correlations in  the Baryon
          Oscillation Spectroscopic Survey (BOSS) 
         \protect\cite{Alam:2017}. The purple point at $z=2.35$ shows $H(z)$ from BAO features in the cross-correlations of
           Ly$\alpha$ absorbers and quasars \protect\citep{Blomqvist:2019}. The blue point at $z=2.34$ shows $H(z)$ from BAO features
             in the correlations of Ly$\alpha$ absorbers \protect\citep{deSainteAgathe:2019}. The magenta point at $z=1.48$ shows $H(z)$ from BAO feaures in  the correlations of quasars \protect\citep{Hou:2020}. The green line shows $H(z)$ for the best-fit base \LCDM\ determined from \Planck\ and the grey bands show $1\sigma$ and $2\sigma$ ranges. The dashed line shows Eq. 
(\ref{equ:phantom}) with parameters chosen to match the SH0ES value of 
$H_0$ at $z =0$.  }

	\label{fig:Hz}

\end{figure}

Figure \ref{fig:Hz} shows various measurements of $H(z)$ from baryon acoustic 
oscillation (BAO) experiments. The normalization
of these measurements assumes the \Planck\ value of the sound horizon
\begin{equation}
 r_d = 147.31 \pm 0.31 \ \Mpc.  \label{equ:sound_horizon}
\end{equation} 
Throughout this paper we will assume that the base \LCDM\ model describes accurately the physics at early times
and so $r_d$ is fixed to Eq. (\ref{equ:sound_horizon}). The sources for the
observational data points are listed in the figure caption. The green line shows $H(z)$ for the best fit base \LCDM\
cosmology determined from \Planck\ and the grey bands show $1\sigma$ and $2 \sigma$ ranges. The green line approaches the value
$H_0^P = 67.44 \Hunit$  asymptotically as $z \rightarrow 0$. As long as $r_d$ remains fixed, apparently the only way to 
reconcile the BAO data with the SH0ES value of $H_0$ is to modify the base \LCDM\ curve. For example,  the dashed line
in Fig. \ref{fig:Hz} shows the relation
\begin{eqnarray}
H(z) &=&  H^f_0  \bigg[ \Omega_m(1+z)^3 +  \qquad \qquad  \qquad \qquad \nonumber  \\
     & & \qquad     (1 - \Omega_m) \left (1+ \Delta\exp(-(z/z_c)^\beta)\right)  \bigg]^{1/2}, \qquad  \label{equ:phantom}
\end{eqnarray}
with parameters $H^f_0 = H^P_0$,  $\Omega_m = 0.31$, $\Delta = 0.30$, $z_c = 0.1$ and $\beta = 2$. With this choice of parameters, the value of
$H_0$ matches the SH0ES value whilst matching the BAO $H(z)$ measurements at $z> 0.3$.  

\footnotetext{The data and covariance matrices are from the file BAO\_consensus\_covtot\_dM\_Hz.txt
  downloaded from \url{http://www.sdss3.org/science/BOSS\_publications.php}}

If  the dashed curve is interpreted as a variation in the equation of state of the dark energy,
then it necessarily requires a phantom equation of state, $w< -1$,  at low redshifts. Alternatively, one might imagine 
that transference of energy between the dark matter and dark energy results in something like the dashed curve.
Models of both  types have been proposed as  `solutions' to the Hubble tension as summarized in  
\cite{diValentino:2021}.  {\it These `solutions' are not viable because the SH0ES team does not directly measure $H_0$.}

In fact, the SH0ES team measure the absolute peak magnitude, $M_B$, of
Type Ia supernovae (SN), assumed to be standard candles,  by calibrating the distances of SN host
galaxies to local geometric distance anchors via the Cepheid period
luminosity relation. The magnitude $M_B$ is then converted into a value of
$H_0$ via the magnitude-redshift relation of the Pantheon SN sample \citep{Scolnic:2017}
of supernovae in the redshift range $0.023 < z < 0.15$. All of the proposed late time `solutions' to
the Hubble tension reviewed in \cite{diValentino:2021} interpret the SH0ES $H_0$ measurement as
a measurement of the  value of  $H(z)$ as $z \rightarrow 0$ (often imposing a SH0ES `$H_0$ prior')
 without investigating whether the `solution'
is consistent with the magnitude-redshift relation of Type Ia SN. It is hardly advancing our
understanding if authors propose solutions to the $H_0$ tension that are 
inconsistent with the measurements that they are trying to explain. 

This point was first made by \cite{Lemos:2019} and more recently by
\cite{Benevento:2020} and by \cite{Camarena:2021}, but has been
comprehensively ignored in recent literature. The purpose of this
paper is to show how theoretical models exploring new physics at late
time should be compared with distance ladder measurements, avoiding
the use of a SH0ES $H_0$ prior. I will adopt a `dynamics free'
approach to this problem and show that late time modifications of the
\LCDM\ expansion history cannot resolve the Hubble tension.

\section{The Inverse Distance Ladder}
\label{sec:inverse}

We will write  the metric of space-time  as 
\begin{equation}
   ds^2 = c^2 dt^2 - R^2(t) (dx^2 + dy^2 +dz^2),  \label{equ:metric1}
\end{equation}
adopting  a spatially flat geometry consistent with the
very tight experimental  constraints on spatial curvature \citep{Efstathiou:2020b}. 
The Hubble parameter, $H = R^{-1} dR/dt$, then fixes the
luminosity distance $D_L(z)$ and comoving angular diameter distance
$D_M(z)$ according to
\begin{equation}
 D_L(z) = c(1+z)\int_0^{z}{dz^\prime \over H(z^\prime)}, \\ 
\quad  D_M(z) = {D_L(z) \over (1+z)}.  \label{equ:metric2}
\end{equation}
Standard candles and standard rulers can therefore be used to constrain
$H(z)$ independently of any dynamics \citep{Heavens:2014, Bernal:2016, Lemos:2019, Aylor:2019} . As long as the relations
of Eq. (\ref{equ:metric2}) are satisfied, it does not matter whether  modifications to  the functional form 
of $H(z)$  are caused by by changes to the equation of state of dark energy or interactions between dark matter
and dark energy. 

\begin{figure*}
	\centering
	\includegraphics[width=85mm, angle=0]{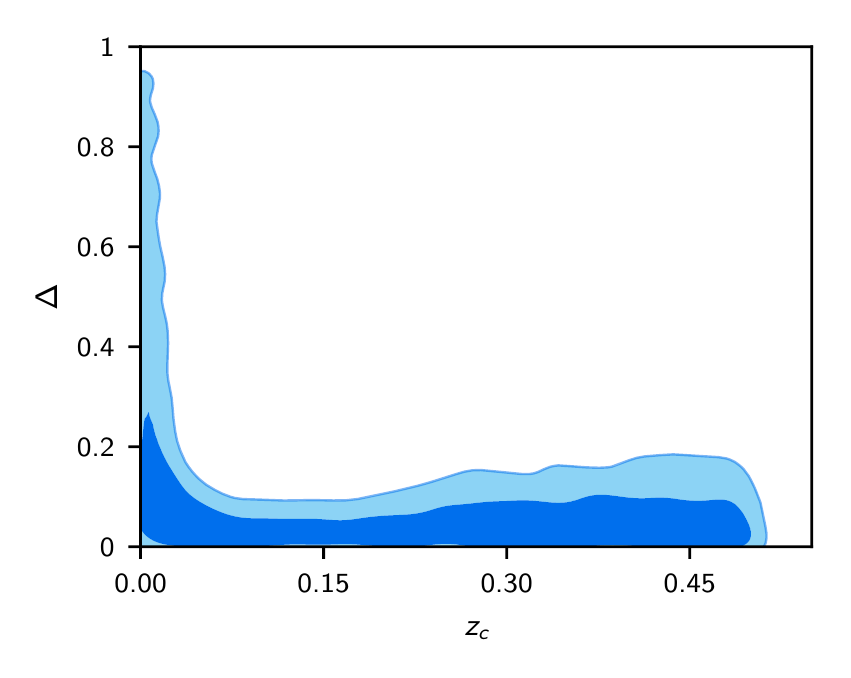} 	\includegraphics[width=85mm, angle=0]{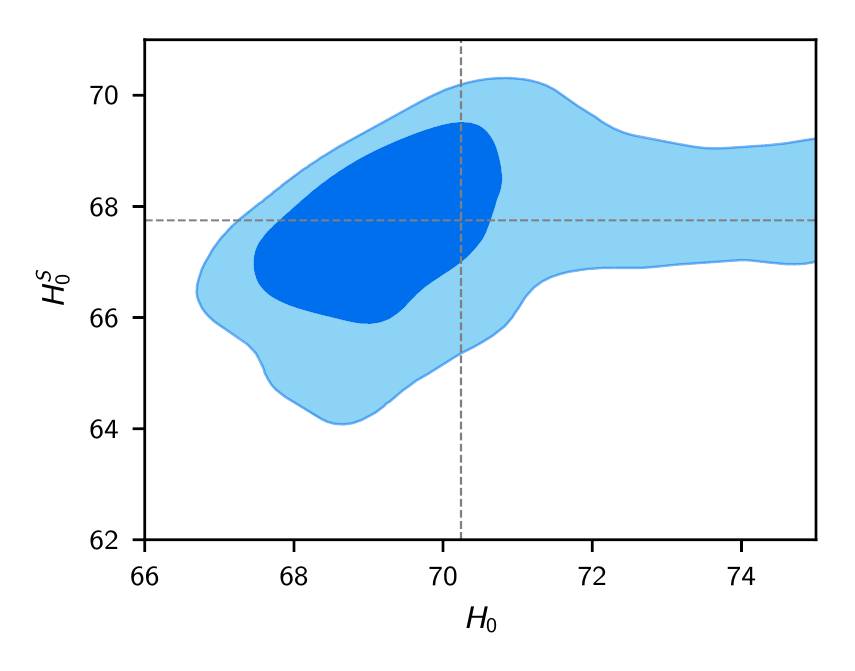} 
	\caption{68 and 95\% constraints on the parameters $\Delta$ and $z_c$ (left hand panel) and the SH0ES-like
         parameter $H^S_0$ of Eq. (\ref{equ:H0Sb}) and $H_0$. (right hand panel). The dashed lines in the right hand panel show the best fit values of $H^S_0$ and $H_0$.  }

	\label {fig:params}

\end{figure*}

A standard candle with absolute magnitude $M$ at redshift $z$ will have
an apparent magnitude
\begin{subequations}
\begin{eqnarray}
m &= & M + 25 + 5 \log_{10} D_L(z),  \label{equ:SH0ES1a}\\
  & = & -5a + 5 \log_{10} c \hat d_L(z) \label{equ:SH0ES1b}
\end{eqnarray}
\end{subequations}
with $D_L$ in units of Mpc.  In  (\ref{equ:SH0ES1b}), $a$ is the intercept of the magnitude-redshift
relation, $5a =-(M + 25 - 5\log_{10} H_0)$ and $\hat d_L(z)  = H_0D_L(z)/c$. The SH0ES Cepheid data allow one to calibrate the
absolute magnitude $M_B$ of Type Ia SN. Combining the geometrical distance estimates
of the maser galaxy NGC 4258 \citep{Reid:2019}, detached eclipsing binaries in the Large Magellanic
Cloud \citep{Pietrzynski:2019} and GAIA early data release 3 (EDR3, \cite{Lindegren:2020a, Lindegren:2020b})  
parallax measurements of 75 Milky Way Cepheids with HST photometry as reported 
in \cite{Riess:2021} (hereafter R21),  the SH0ES Cepheid photometry and Pantheon SN peak magnitudes give
\begin{equation}
M_B = -19.214 \pm \  0.037  \ {\rm mag.}  \label{equ:SH0ES2a}
\end{equation}
(see Sect. \ref{sec:global_fits}). 
To estimate $H_0$, R16 determine the intercept of the Pantheon SN magnitude-redshift relation by 
fitting the low redshift expansion to the luminosity distance
\begin{equation}
\hat d_l(z) = z\left [ 1 + (1 - q_0) {z \over 2} - {1 \over 6}(1 - q_0 - 3 q_0^2 + j_0)z^2 \right ], \label{equ:SH0ES2}
\end{equation}
over the redshift range $z = 0.023$ to $z =0.15$,  with the deceleration and jerk parameters set to $q_0=-0.55$ and $j_0 = 1$ (close to the values for base \LCDM,  $q_0 = -0.535$, $j_0 = 1$). They find
\begin{equation}
a_B = 0.71273 \pm 0.00176,  \label{equ:SH0ES3}
\end{equation}
which, together with Eq. (\ref{equ:SH0ES2a}), gives
\begin{equation}
H_0 = 74.1 \pm 1.3 \Hunit,  \label{equ:H0}
\end{equation}
(see Table \ref{tab:fits}). This is slightly higher than the value $H_0 = 73.2 \pm 1.3 \Hunit$ quoted in 
 R21 reflecting differences in the period ranges and photometric samples used in my analysis. These differences are
unimportant for this paper.  (Note, however,  that   the global fits to the SH0ES Cepheid data  using the GAIA EDR3 parallaxes have  
puzzling features as discussed in Sect. \ref{sec:global_fits}.)

To apply the inverse distance ladder, I  follow closely the analysis described in \cite{Lemos:2019}. $H(z)$ is parameterized
by Eq. (\ref{equ:phantom}) and the parameters of the model are determined by fitting to Pantheon SN magnitudes  and 
the BAO $D_M(z)$ and $H(z)$
measurements from the references given in the caption to Fig. \ref{fig:Hz}, supplemented by the
$D_V(z) = (D_M^2(z) cz/H(z))^{1/3}$  measurement at $z=0.106$ from \cite{Beutler:2011}. 
The free parameters of the model 
are $H_0^f$, $\Omega_m$, $\Delta$, $z_c$, $\beta$  and $M_B$ with uniform priors 
as listed in Table \ref{tab:parameters}. To compare with the  BAO results, I adopt
 a Gaussian prior on the sound horizon $r_d$ with the parameters of Eq. (\ref{equ:sound_horizon}). 
I use the {\tt MULTINEST} sampling algorithm \citep{Feroz:2009, Feroz:2011} to explore the parameter
space.

\begin{table}

\begin{center}

\caption{Results of applying the inverse distance ladder. The table lists the mean values of the 
parameters and their $1\sigma$ error. The last column lists the ranges over which a uniform prior
is applied to the parameters.   The parameters $H_0$ and $H^S_0$ are derived parameters (see Eqs.
\ref{equ:phantom} and  \ref{equ:H0Sb}). The units of $H^f_0$, $H_0$ and $H^S_0$ are $\Hunit$.}

\label{tab:parameters}

\smallskip

\begin{tabular}{l|c|c|} \hline 
parameter  &  fit   & prior range    \\ \hline
$H^f_0$    &   $68.13 \pm 1.00$ &  $60 \ \textendash \ 80$  \\
$\Omega_m $    &   $0.306 \pm 0.017$ &  $0.25 \ \textendash \ 0.35$  \\
$\Delta $    &   $0.107 \pm 0.162$ &  $0.0 \ \textendash \ 1.0$  \\
$z_c $    &   $0.167 \pm 0.091$ &  $0.001 \ \textendash \ 0.5$  \\
$\beta $    &   $2.45 \pm 0.86$ &  $1.0 \ \textendash \ 4.0$  \\
$M_B$    &   $-19.387 \pm 0.021$ &  $-19.0 \ \textendash \ -19.5$  \\
$H_0$    &   $70.5 \pm 3.6$ &    \textendash \\
$H^S_0$    &   $67.73 \pm 0.97$ &  \textendash \\  \hline

\end{tabular}
\end{center}
\end{table}

\begin{figure*}
	\centering
	\includegraphics[width=58mm, angle=0]{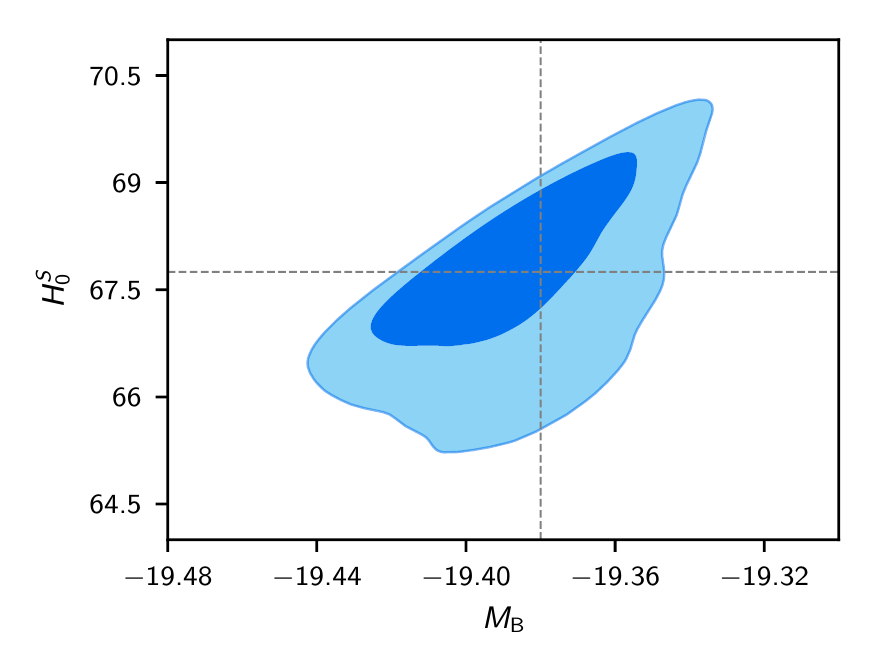} 	\includegraphics[width=58mm, angle=0]{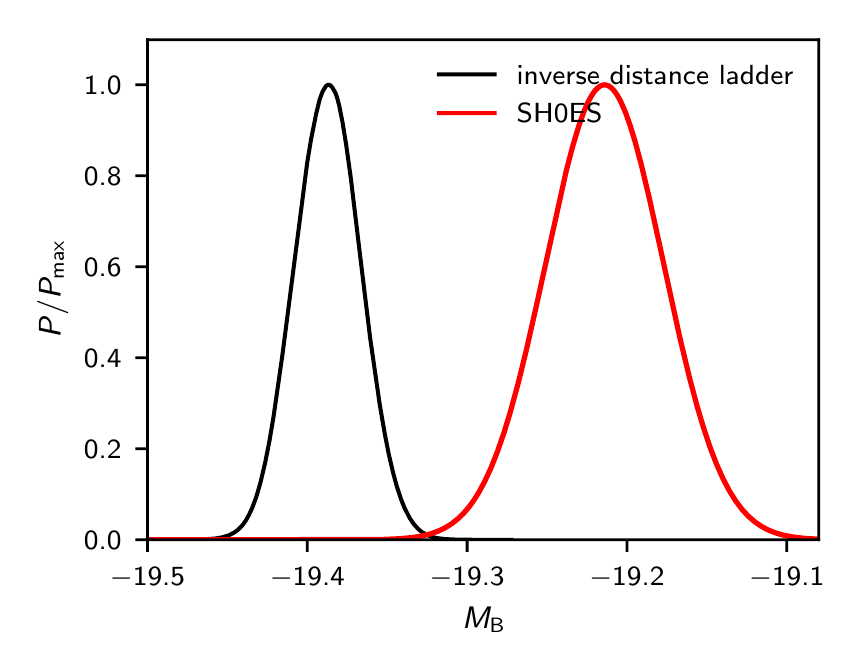}  	\includegraphics[width=58mm, angle=0]{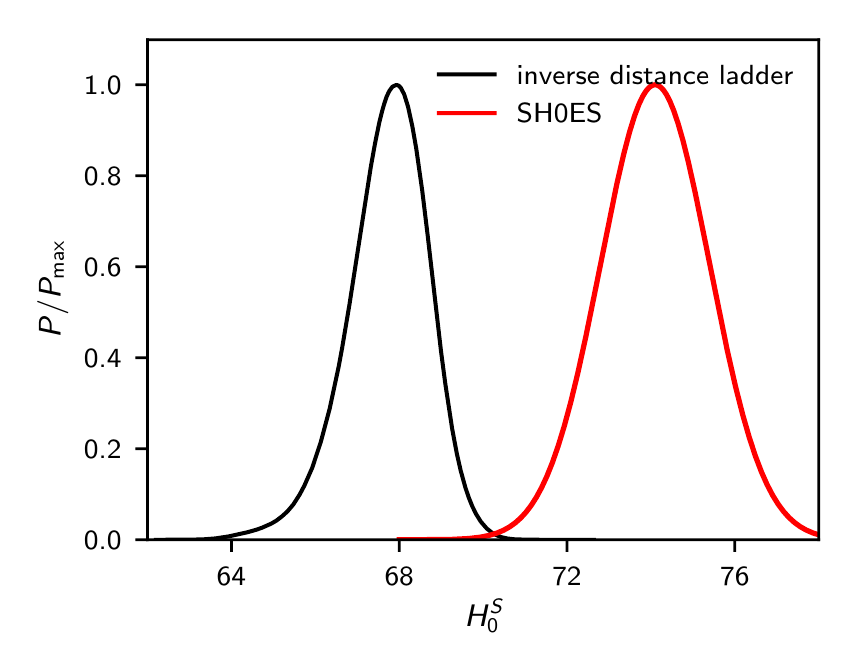} 

	\caption{The left hand panel shows 68 and 95\% constraints on the parameters $H_0^S$ and $M_B$. The dotted lines
show the mean values of these parameters listed in Table \ref{tab:parameters}. The midle panel shows the marginalised posterior distributions
of the SN peak absolute magnitude $M_B$ determined from the inverse distance ladder discussed in this paper (black line)
compared with the posterior distribution of $M_B$ determined from the SH0ES data (red line). The right hand panel shows
the equivalent plot, but for the parameter $H^S_0$ instead of $M_B$.}

	\label {fig:params2}

\end{figure*}

The constraints on these parameters are summarized in Table \ref{tab:parameters}.  The left hand plot in Fig. \ref{fig:params} shows the $1\sigma$
and $2\sigma$ constraints on the parameters $\Delta$ and $z_c$\footnote{ Evidently, the parameter $z_c$ runs into the upper range of its prior, but this is unimportant.}. The key point here is that the parameter $\Delta$ is well constrained for values of $z_c \simgt 0.05$ because at these redshifts the form of $H(z)$  is tightly constrained by the SN magnitude-redshift relation. At lower values of $z_c$,  the parameter $\Delta$ becomes poorly constrained by the SN
magnitude-redshift relation and solutions  with high values of $H_0$ are allowed. This reinforces the conclusions of \cite{Benevento:2020} and \cite{Camarena:2021} that the SN data are insensitive to changes in the dark energy equation of state at very late times\footnote{It is worth mentioning that the SN host galaxies of R16 and \cite{Freedman:2019} are very nearby, with redshifts $z \simlt 0.007$. Yet as shown in Fig. 7 of \cite{Freedman:2019}, their velocity flow corrected distances define a Hubble diagram with very little  scatter.  
There is therefore no evidence for an abrupt change to the equation of state at very low redshifts.}. 

We can compute a derived quantity $H^S_0$ that is equivalent to the SH0ES estimate of $H_0$ 
\begin{subequations}
\begin{eqnarray}
\hspace{-0.15in}a_B &\hspace{-0.1in}=& \hspace{-0.1in} \left (\sum_{ij} C^{-1}_{ij} (\log_{10}\hat d_L(z) - 0.2 m_B(i)) \right ) / \sum_{ij} C^{-1}_{ij} \qquad\qquad    \label{equ:H0Sa} \\
H^S_0 &\hspace{-0.1in}=& \hspace{-0.1in}10^{0.2(M_B + 5 a_B + 25)}  \label{equ:H0Sb}
\end{eqnarray}
where $C$ is the covariance matrix of the Pantheon SN magnitudes and the sums in Eq. (\ref{equ:H0Sa}) extend all SN in the Pantheon sample with redshifts in the range $0.023 - 0.15$. 
\end{subequations}

The right hand plot in Fig. \ref{fig:params} shows $H^S_0$ plotted
against the true value of $H_0$. One can see the long tail extending
to high values of $H_0$. These high values arise in solutions with
$z_c \simlt 0.05$ and high values of $\Delta$ corresponding to
phantom-like equations of state. However, the SH0ES analysis is
oblivious to these high values of $H_0$. Instead, a SH0ES type
analysis would infer a value close to the estimate $H^S_0$ of
Eq. (\ref{equ:H0Sb}),  which is always low. For these models $H^S_0 =
67.75 \pm 1.01 \Hunit$, discrepant with Eq. (\ref{equ:H0})
by $3.9 \sigma$ {\it despite the ability of the model to mimic extreme
  phantom-like equations of state.} It is also worth noting that the parameters
$\Omega_m$ and $H_0^S$ are each within about $0.4\sigma$ of the base \LCDM\ values
inferred from \Planck. There is not even a hint from these data for any phantom-like
physics.

The discrepancy between the inverse distance ladder and the SH0ES data is illustrated clearly in Fig. \ref{fig:params2}. 
The left hand panel shows the $1\sigma$ and $2 \sigma$ constraints on $H^S_0$ and $M_B$. The central panel shows the
posterior distribution of $M_B$ determined from the inverse distance ladder (black line) compared with the result
of Eq. (\ref{equ:SH0ES2}) derived from  the SH0ES data  (red line). The right hand panel shows the posterior distribution
of $H^S_0$ compared with Eq. (\ref{equ:H0}). The SH0ES results are clearly discrepant with the inverse distance ladder.
As long as the \Planck\ value of $r_d$ is correct and the relations of Eq. (\ref{equ:metric2}) apply,  {\it
 modifications to  the  expansion history at late times cannot explain the SH0ES data.}

\section{Should one use a SH0ES $H_0$ prior?}
\label{sec:global_fits}

The first draft of this paper generated a large number of
comments. Many of the negative comments cited the paper by
\cite{Dhawan:2020} which explored a number of parametric forms for the
expansion history (interpreted as modifications to the dark energy
equation of state) constrained to fit the Pantheon magnitude-redshift
relation.  \cite{Dhawan:2020} concluded that the distance ladder
values of $H_0$ inferred for these expansion histories agreed to
within about $0.5 \Hunit$ (excluding
 models with a sharp transition in the equation of
  state at very low redshift, which were not considered by
  \cite{Dhawan:2020}).  Critics of my paper have argued that the
insensitivity of $H_0$ to variations in the expansion history found by
\cite{Dhawan:2020} means that it is safe to use a SH0ES $H_0$ prior in
cosmological parameter analysis {\it even if the Pantheon sample is excluded from
the analysis}. This is incorrect.

The key  point is that the Pantheon SN sample is an essential part of 
the SH0ES distance ladder measurement of $H_0$. Summarising the entire SH0ES
analysis in terms of one parameter, namely the value of $H_0$,  represents a 
huge and lossy compression of the Cepheid+SN data. In particular, all information on the 
shape of the SN magnitude-redshift relation is lost. If one then imposes a SH0ES
$H_0$ prior but ignores the Pantheon SN data, it is possible to infer 
evidence for phantom like dark energy as illustrated by the dotted line in 
Fig. \ref{fig:Hz}. However, such a solution is strongly disfavoured by the 
Pantheon magnitude-redshift relation.

If one wants to investigate consequences of new late-time physics, a rigorous way of
compare with the SH0ES results, as noted by \cite{Camarena:2021},  is to drop $H_0$ as a parameter in 
favour of the SN peak absolute magnitude $M_B$. In other words,  rather than  explaining the `Hubble tension' one should 
instead focus on the `supernova absolute magnitude tension' since this is what the Cepheid calibrations are designed
to measure. The goal then is to  find a late time solution that
brings $M_B$ measured from distant supernovae into agreement with the value inferred from Cepheid  measurements.
This necessarily involves analysing a uniformly calibrated supernova sample such as 
Pantheon\footnote{Similar remarks apply to the tip of the red giant branch distance ladder \cite{Freedman:2019},
but with the Carnegie Supernova Project \citep{Hamuy:2006, Krisciunas:2017} replacing the Pantheon sample.}. 
The following procedure is statistically rigorous:

\begin{figure}
	\centering
	\includegraphics[width=85mm, angle=0]{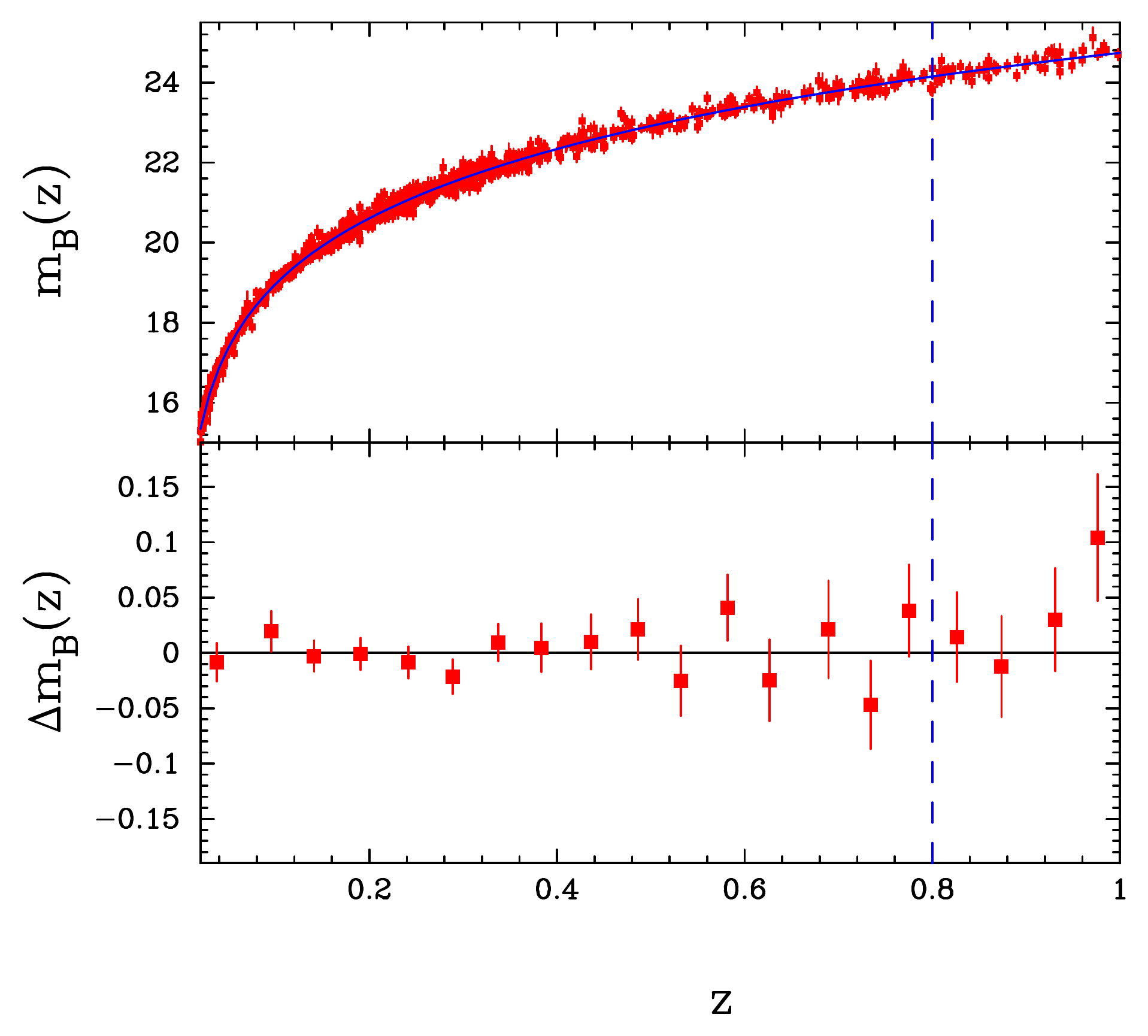} 
	\caption{The upper panel shows the magnitude-redshift relation for the Pantheon sample, together with the best
fit (solid line)  assuming the expansion history of Eq. (\ref{equ:log}). The vertical dashed line shows the maximum redshift used in the fit. The lower panel shows maximum likelihood band averaged residuals with respect to the best fit,
together with $1\sigma$ errors.}

	\label{fig:pantheon}

\end{figure}

\smallskip 

\noindent
[1] If correlations between the host\footnote{We are using the term
  `host' as shorthand to denote  galaxies with Cepheid distance
  moduli that hosted a Type 1a SN. } galaxy SN magnitudes and the
magnitudes of more distant supernovae are ignored (as in the the SH0ES
papers and this paper)  the SH0ES Cepheid measurements can be
summarized by the posterior distribution of the SN peak absolute
magnitude $M^1_B$.  If one wants to take into account 
correlations between the magnitudes of host and distant SN, the SH0ES data must be
summarized in terms of a vector of distance moduli ${\bs \mu}$ and an
associated covariance matrix as described in Appendix
\ref{sec:appendix}.

\smallskip

\noindent
[2] To test a theoretical model, carry as parameters an absolute
magnitude $M^1_B$ for the Cepheid SN host galaxies and an absolute
magnitude $M^2_B$ for the more distant SN in the Pantheon sample (taking into account correlations
with the Cepheid SN magnitudes, if necessary).  If
the posteriors of $M^1_B$ and $M^2_B$ overlap, then one has a
candidate solution to the `supernova absolute magnitude tension'.

\smallskip

\noindent
[3] If there is substantial overlap between the posteriors of $M^1_B$
and $M^2_B$, one can replace these parameters by a single parameter
$M_B$. The best fit value of $M_B$ for a chosen cosmology is 
equivalent to a best fit value of $H_0$.

\begin{figure}
	\centering
	\includegraphics[width=85mm, angle=0]{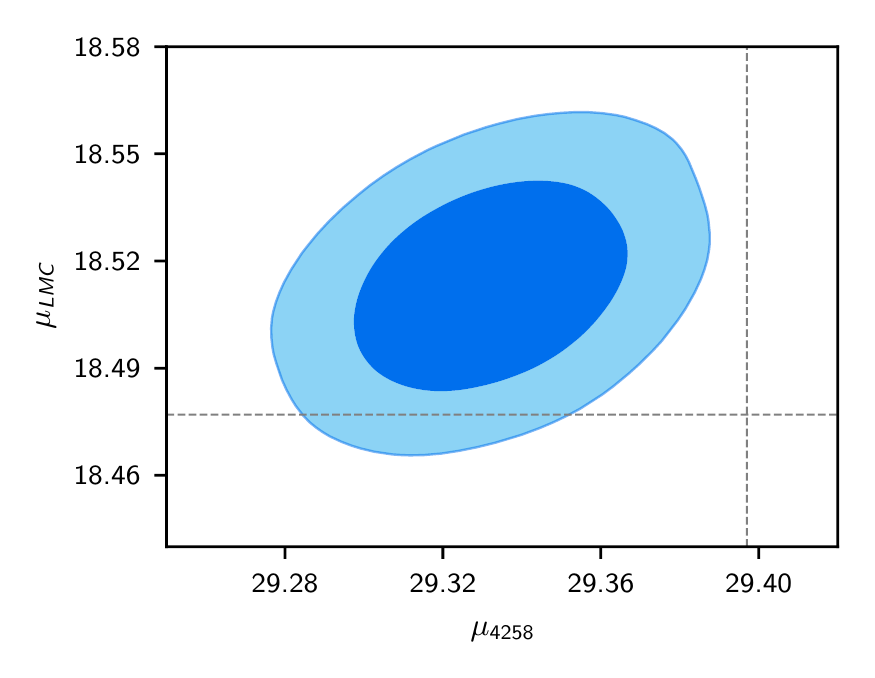} 
	\caption{68\% and 95\% constraints on the NGC 4258 and LMC distance moduli for the three anchor global fit
summarized in column 5 of Table \ref{tab:fits}. The best fit geometrical distance moduli of \citep{Reid:2019} and
\citep{Pietrzynski:2019} (which are included as priors in the global fit) are shown by the dotted lines.}

	\label{fig:anchors}

\end{figure}

\begin{table*}

\begin{center}

\caption{Determination of $H_0$ from three geometrical  distance anchors 
(NGC 4258 maser distance, LMC detached eclipsing binaries and GAIA DR3 parallaxes
to Milky Way Cepheids) using different fits to the expansion history. Column 2 gives results 
assuming a Gaussian prior on $a_B$ of $0.71273 \pm 0.00176$ as in R16 and subsequent SH0ES
papers. Columns 3 and 4 give results of fitting the $q_0, j_0$ expansion history of Eq. (\ref{equ:SH0ES2})
to the  Pantheon sample for SN in the redshift range $0.023$ to $z_{\rm max}$. Column 5 shows the results of 
fitting Eq. (\ref{equ:log}) to the Pantheon sample over the redshift range  $0.023-0.8$. The accuracy of
the best fit to the Pantheon magnitude-redshift relation in this case is illustrated in Fig. \ref{fig:pantheon}. 
The units of the parameter $zp$ is $\mu$arcsec and $H_0$ is  in units of $\Hunit$.}

\label{tab:fits}

\smallskip

\begin{tabular}{l|c|c|c|c} \hline 
parameter  &  $a_B$ prior    &   Eq. (\ref{equ:SH0ES2}) ($z_{\rm max} = 0.50$) &  Eq. (\ref{equ:SH0ES2})
 ($z_{\rm max} = 0.80$) &  Eq. (\ref{equ:log}) ($z_{\rm max} = 0.80$) \\ \hline
$M^W_H$ &     $-5.922 \pm 0.017$  & $-5.922 \pm 0.017$ & $-5.922 \pm 0.017$ &  $-5.922 \pm 0.017$ \\
$b$    &   $-3.24 \pm 0.02$ &     $-3.24 \pm 0.02$ & $-3.24 \pm 0.02$ & $-3.24 \pm 0.02$\\
$Z_w $    &   $-0.22 \pm 0.06$  & $-0.22 \pm 0.05$  &$-0.22 \pm 0.05$  &  $-0.22 \pm 0.06$ \\
$zp$    &   $-17.1 \pm 4.8$  &  $-17.2 \pm 4.7$  &$-17.1 \pm 4.7$ & $-16.8 \pm 4.8$ \\
$M_B$    &   $-19.214 \pm 0.037$   & $-19.213 \pm 0.036$ & $-19.212 \pm 0.037$ & $-19.209 \pm 0.037$ \\
$a_B$    &   $0.71273\pm 0.00176$  & $0.7174 \pm 0.0028$ & $0.7164 \pm 0.0026$ & $0.7162 \pm 0.0033$ \\
$H_0$    &   $74.1 \pm 1.3$  & $75.0 \pm 1.3$ & $74.8 \pm 1.3$ & $74.9 \pm 1.4$ \\  \hline

\end{tabular}
\end{center}
\end{table*}

\smallskip

In the above procedure, there
is no danger of reaching erroneous conclusions on late
time physics through the innapproprate use of an $H_0$ prior. In addition, the
Pantheon data is only used only once in testing a particular theoretical model 
\citep[c.f.][]{Camarena:2021}. Furthermore, this approach remains valid for models
in which the expansion history changes at $z \simlt 0.05$.

What procedure should forward distance ladder measurements follow in
reporting a value for $H_0$?   It is clear that one should fit the expansion history using the
Pantheon sample as part of the $H_0$ analysis rather than adopting fixed values for $q_0$ and $j_0$
as in R16.
To illustrate this, Table \ref{tab:fits} shows results of global fits to the
 SH0ES Cepheid  and Pantheon SN magnitude-redshift relation combining three geometrical distance anchors, 
as described in Sect. \ref{sec:inverse}. The Cepheid period luminosity is
fitted to  (see e.g. R21) 
\begin{equation}
m^i_j =  \mu_i + M^W_H + b (\log_{10} P_j - 1) + Z_w \Delta [O/H]_j  , \label{equ:PL}
\end{equation}
where $m^i_j$ is the H-band Weisenheit apparent magnitude of Cepheid
$j$ in galaxy $i$, $P_j$ is the period of Cepheid $j$ in units of days
and $\Delta [O/H]_j$ is the metallicity assigned to Cepheid $j$
relative to Solar metallicity. As in R21 I include a constant offset
$zp$ to the GAIA EDR3 parallaxes as a free parameter.  Column 2 shows
results assuming the R16 prior on $a_B$ (Eq. \ref{equ:SH0ES3}) which
gives the value of $H_0$ quoted in Eq. (\ref{equ:H0}). The next two
columns show results using the ($q_0$, $j_0$) expansion of
Eq. (\ref{equ:SH0ES2}) which we fit to the Pantheon magnitude-redshift
relation over the redshift ranges $0.023-0.5$ (column 3) and
$0.023-0.8$ (column 4).  The Cepheid period-luminosity parameters, and
the parameter $M_B$ are stable across the columns, as expected since
these parameters are statistically decoupled from the Pantheon
magnitude-redshift relation. However, the value of the Hubble constant
increases by $\sim 0.7 \Hunit$ between columns 2 and 4.

To test the sensitivity of these results to the assumed  expansion history, column 5 in Table \ref{tab:fits}
shows results for the flexible parametric form:
\begin{equation}
H(z) = H_0^f \left[A(1+z)^3 + B + C z + D \ {\rm ln}(1+z)\right]^{1/2}, \label{equ:log}
\end{equation}
with $A$, $B$, $C$ and $D$ as free parameters \citep{Lemos:2019}. The best fit SN magnitude-redshift relation
is shown in Fig. \ref{fig:pantheon}. Equation (\ref{equ:log}) provides a very accurate  fit to the Pantheon 
magnitude-redshift relation and is very close to the relation for the best-fit Planck base \LCDM\ cosmology (as can be seen from Fig. 13
of \cite{Params:2018}). For most purposes, it will be sufficient to  quote a value of $H_0$ based on fits of a  flexible fitting
function (or  a Gaussian process, see e.g. \cite{Shafieloo:2012, Joudaki:2018}) to the Pantheon 
magnitude-redshift relation. However, there remains an uncertainty in this type of forward estimation of $H_0$, which is
difficult to quantify,   if the expansion history deviates from the base \LCDM\ cosmology at redshifts $z \simlt 0.05$
since such models are poorly constrained by the Pantheon sample (cf. Fig. \ref{fig:params}).

It is worth mentioning a peculiar aspect of the solutions in Table
\ref{tab:fits}. \cite{Efstathiou:2020} pointed out a $\sim 3 \sigma$
tension between the NGC 4258 and LMC geometric distance anchors in the
global fits using the R16 Cepheids. This tension becomes stronger
if we include the GAIA EDR3 Cepheid parallaxes. This is illustrated in
Fig. \ref{fig:anchors}, which shows the constraints on the LMC and
NGC 4258 distance moduli ($\mu_{\rm  LMC}$ and $\mu_{4258}$) derived from 
the fit in column 5 of Table \ref{tab:fits}. The  GAIA EDR3 and LMC 
anchors pull the solution towards high values of $H_0$, while the
NGC 4258 maser anchor wants to pull the solution to lower values of $H_0$.
The  three-anchor solutions listed in Table \ref{tab:fits} (and in R21)
are therefore statistically inconsistent.

\section{Conclusions and Discussion}
\label{sec:conclusions}

The Hubble tension has led to a large literature in the last few
years.  Authors of proposed late time solutions to the Hubble tension
have often imposed a SH0ES $H_0$ prior on the Hubble parameter at
$z=0$, leading to erroneous claims of evidence for phantom dark
energy, dark matter-dark energy interactions, or other exotic
late-time physics. The review article by \cite{diValentino:2021} cites
many such examples.

If one wants to investigate consequences of new late-time physics, the
simplest way to compare with the SH0ES results is to drop $H_0$ as a
parameter in favour of the SN peak absolute magnitude $M_B$,
i.e. rather than explaining the `Hubble tension' one should instead
focus on the `supernova absolute magnitude tension'. The goal then is
to find a late time solution that brings $M_B$ into agreement with the
SH0ES measurement. This necessarily involves analysing the Pantheon SN
sample\footnote{Similar remarks apply to the tip of the red giant
  branch distance ladder \cite{Freedman:2019}, but with the Carnegie
  Supernova Project \citep{Hamuy:2006, Krisciunas:2017} replacing the
  Pantheon sample.}.  If one wants to combine the SH0ES data with
other astrophysical data to constrain late time physics, then one
should impose a SH0ES prior on the parameter $M_B$ (or Cepheid calibrated 
distance moduli of SN host galaxies, see Appendix \ref{sec:appendix})  and not on the
parameter $H_0$. 

However, using the Pantheon and BAO data, the inverse distance ladder places
very strong constraints on new physics at late times. The results of
Table \ref{tab:parameters}  show that the data are in excellent agreement with the base
\LCDM\ cosmology determined from \Planck. BAO is now
a  mature field employing analysis techniques that have been 
tested extensively against simulations. There is no good reason to 
ignore these measurements. Neither is there a good reason to ignore the
Pantheon SN sample, since this is an essential part of the SH0ES distance
ladder. It is, therefore, unlikely that changes to the late time expansion history
can resolve the `Hubble tension'. This conclusion is independent of any dynamics,
and independent of perturbations insofar  as the \Planck\ value of $r_d$ is
unaltered.

\section*{Acknowledgements} 

I thank Sunny Vagnozzi for his comments on a draft of this paper. I am grateful to the
many people who have sent me comments on this paper. In particular, I thank Adam Riess,
Dan Scolnic and Pablo Lemos for correspondence on the material discussed in Section 3.

\section*{Data Availability} 

No new data were generated or analysed in support of this research.

\appendix
\section{Including correlations between Cepheid-calibrated SN magnitudes and magnitudes of more distant SN}
\label{sec:appendix}

We denote the covariance matrix of the SN magnitudes as ${\bs M}$ and the covariance matrix of
the  host galaxy distance moduli as ${\bs C}$. Let  $t$
denote the index of the $N_H$ host SN  and $q$  denote the index of the $N_P$ more distant SN in the Pantheon sample that
will link the Cepheid measurements to the
Hubble flow. The predicted peak SN magnitude in a host galaxy is 
\begin{subequations}
\begin{equation}
       m^P_t  =   \mu_t + M^1_B , \label{equ:app1a}
\end{equation}
where $\mu_t$ is the distance modulus to galaxy $t$. The predicted 
peak SN magnitude for a distant SN at redshift $z_q$ is
\begin{equation}
       m^P_q  =   \mu(z_q) + M^2_B , \label{equ:app1b}
\end{equation}
\end{subequations}
where the distance modulus $\mu(z_q)$ is fixed by an assumed cosmology and value of
$H_0$.   We would expect $M^1_B = M^2_B = M_B$ if Type 1a supernovae are standard candles.
Let ${\bs x} = ({\bs m} -{\bs m^P})$, where ${\bs m}$ is the data vector of SN peak magnitudes, 
and partition the vector ${\bs  x}$ as (${\bs y}$, $\bs w$)
where ${\bs y}$ describes the  hosts and ${\bs w}$ describes the more distant SN.
Let us partition the covariance matrix ${\bs M}$ as follows:
\begin{equation}
 {\bs M}^{-1} = \left ( \begin{array} {c|c} 
  {\bs D} &  {\bs E}  \\  \hline
  {\bs E}^T &   {\bs F}
              \end{array} \right ), \label{equ:app2}
\end{equation}
where ${\bs D}$, ${\bs E}$ and ${\bs F}$ have dimensions  $N_H\times N_H$, $N_H\times N_P$ an
$N_P \times N_P$ respectively.

If correlations between host and distant SN magnitudes are negligible,
the Cepheid analysis is decoupled from the analysis of the distant SN in the Pantheon
sample. In this case, the Cepheid analysis can be summarized by the
posterior distribution of $M^1_B$ and the likelihood of the distant SN is\footnote{All
  likelihoods in this section are defined to within an additive
  constant.}
\begin{equation}
-2\ln {\cal L} =   {\bs w}^T {\bs F} {\bs w}.   \label{equ:app4}
\end{equation}
Given an assumed functional form for $D_L(z)$, Eq. (\ref{equ:app4}) can be used to determine a probability distribution for
 $H_0$ given a distribution for $M^1_B = M^2_B = M_B$ . This
is the approximation adopted by the SH0ES team and in the main body of this paper.

To take into account correlations between host and distant SN magnitudes, a more complex procedure
is necessary. In this case, the Cepheid analysis can no longer be summarized by the posterior distribution of $M^1_B$.
Instead the Cepheid analysis needs to be summarized in terms of
the distance moduli for the SN hosts and a covariance matrix. It is a very good
approximation to assume a Gaussian probability distribution 
\begin{equation}
-2\ln {\cal L} =   ({\bs \mu} - {\bs \mu}^m)^T {\bs C}^{-1} ({\bs \mu} - {\bs \mu}^m),  \label{equ:app5}
\end{equation}
where ${\bs \mu}^m$ are the mean values of the host distance moduli determined from the MCMC chains
and ${\bs C}$ is their covariance matrix  

There is, however, no need to carry
the $N_H$ components of  ${\bs \mu}$ as parameters, since it is possible to integrate over
these variables. After some algebra, one can show that the likelihood can be written as 
\begin{eqnarray}
-2\ln {\cal L} &=&   {\bs x}^T {\bs M}^{-1} {\bs x} \qquad \qquad \qquad \qquad   \nonumber \\
 -   & & \hspace{-0.3in}   [{\bs D}{\bs y} + {\bs E}{\bs w}]^T ({\bs C}^{-1} + {\bs D})^{-1}     [{\bs D}{\bs y} + {\bs E}{\bs w}], 
 \label{equ:app6}
\end{eqnarray}
where the  $\mu_t$ in Eq. (\ref{equ:app1a})  are  replaced by the maximum likelihood values
$\mu^m_t$. 

In the limit that correlations between host and distant magnitudes can be neglected, the host 
component of the likelihood
becomes
\begin{equation}
-2\ln {\cal L} =   {\bs y}^T {\bs D} [ {\bs I} - ({\bs C}^{-1} + {\bs D})^{-1}  {\bs D}] {\bs y}.  \label{equ:app7}
\end{equation}
Equation (\ref{equ:app7}) reproduces the results of Table 2 to within the precision of the MCMC chains.

\bibliographystyle{mnras}
\bibliography{toH0} 
\end{document}